# Investigation of Thermoelectric properties of Magnetic Insulator FeRuTiSi Using First Principle Calculation

Saurabh Singh[1, a)] Shubham Singh,[2)] Nitinkumar Bijewar,[2)] and Ashish Kumar [3,b)]

[1](*Toyota Technological Institute, Hisakata, 2-12-1, Tempaku-ku, Nagoya 468-8511, Japan.*)
[2](*Department of Physics, university of Mumbai, kalina Campus, Santacruz (E)-400098, Mumbai, India.*)
[3](*Inter-University Accelerator Centre, Aruna Asaf Ali marg, Vasant Kunj, New Delhi-110067, India.*)

[a)]Corresponding author: saurabhsingh@toyota-ti.ac.jp
[b)]Corresponding author: ashish@iuac.res.in

**Abstract.** In this work, we have investigated the electronic structure and thermoelectric properties of quaternary heusler alloy, FeRuTiSi, using first principle DFT tools implemented in WIEN2k and BoltzTraP code. Electronic structure calculations using TB-mBJ potential shows appearance of flat band at the conduction band edge, thus electron in conduction band have the large effective mass ($m_e^*$), and therefore mainly contribute for negatively large value of Seebeck coefficient (S). This alloy has indirect band gap of 0.59 eV, and shows the *n*-type transport behavior. Under the constant relaxation time approximation ($\tau = 10^{-14}$ s), temperature dependent Seebeck coefficient, electrical conductivity ($\sigma$), and electronic thermal conductivity ($\kappa_e$) were also estimated. The maximum *figure-of-merit (ZT)*, for the FeRuTiSi compound is found to be ~0.86 at 840 K, with n-type doping, which suggests that this quaternary alloy can be a good candidate among the *n*-type material for thermoelectric applications in high-temperature region.

## INTRODUCTION

Quaternary heusler alloy (QHA) with half-metallic electronic structure have been extensively investigated in past several years for the spintronic applications, as they exhibit high spin-polarization property [1,2]. In the simple way, one can define the crystal structure such as ABCD with 1:1:1:1 stoichiometry. The crystallographic structure for prototype QHA is LiMgPdSn, classified by the space group no 216: *F-43m* [3]. The systematic study on various QHA have been carried out both experimentally and theoretically [1,2]. The majority of the investigation have been focused on magnetic material for spin driven property originated from the magnetic constituent of the alloy. Moreover, depending on the number of valance charge present in the unit cell, some of the QHA possess the insulating electronic structure with non-magnetic ground state consistent with Slater-Pauling rule (M = Nv – 24) [4]. These insulating QHA with narrow band gap can be the choice for thermoelectric applications, as mankind need demands an alternate source of energy for the future.

Thermoelectric (TE) materials have the ability to convert waste heat into useful electrical energy. For better efficiency, TE material should have the high *figure-of-merit (ZT)* defined as ZT = ($\alpha^2\sigma$)T/$\kappa$, where $\alpha$, $\sigma$, $\kappa$ and T are Seebeck coefficient, electrical conductivity, electronic thermal conductivity and absolute temperature, respectively [5]. Among various QHA, recent electronic calculations reported on FeRuTiSi shows that this alloy have band gap of 0.4 eV [6]. But, the calculations are limited upto the electronic structure by using the ultrasoft pseudopotential with a plane-wave basis set method implemented in CASTEP code. In order to check the potential capability of this material for TE application, an investigation on temperature dependent TE coefficients is required. Thus, we performed a systematic study on this material using first principle calculations and check the TE property in a detailed way. Using TB-mBJ potential, we found the indirect band gap ~0.59 eV. The transport properties calculations result using BoltzTraP code provide *ZT* = 0.86 at 840 K temperature for *n*-type FeRuTiSi, which is very good to consider this material for thermoelectric application.

# COPUTATIONAL DETAILS

For the structural optimization and electronic structure calculations, we have used density functional theory based on *full potential linearized augmented plane-wave* (FP-LAPW) method, implemented in WIEN2K code [7]. The exchange correlation functional of generalized gradient approximation (GGA) formulated by Perdew et al. along with modified Becke-Johnson potential (TB-MbJ) given by Tran and Blaha, were used for calculations of electronic band structure [8,9]. In unit cell, FeRuTiSi occupy the Wyckoff positions similar to LiMgPdSn structure which are such as Fe(1/4,1/4,1/4), Ru(3/4,3/4,3/4), Ti(1/2,1/2,1/2), and Si(0,0,0). The k-mesh of size 10×10×10 was used for geometrical optimization, whereas highly dense k-mesh values were considered for accurate calculation of electronic density of states (20×20×20), and transport properties (50×50×50). The convergence criteria for self-consistency field calculation was set to the $10^{-5}$ Ry. To investigate the temperature dependent thermoelectric properties, transport coefficients calculations were performed using Semi-classical Boltzmann transport theory implemented in the BoltzTraP code [10].

## REULTS AND DISCUSSIONS

Figure 1a shows the unit cell of the FeRuTiSi. In order to get the equilibrium lattice parameter, the structural optimization using PBE exchange correlation functional is carried out, and obtained results is shown in Fig 1b. Equation of states fit by using Birch-Murnaghan expression gives the equilibrium lattice parameter of 5.871 Å, and bulk-modulus value of ~228 GPa. Spin-polarized calculation shows the non-magnetic ground state.

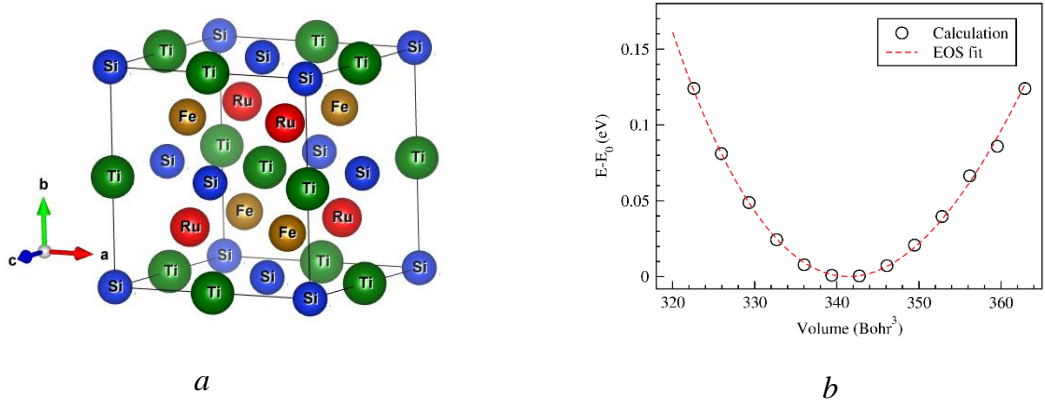

*a*                *b*

Fig 1. (*a*) Unit cell of FeRuTiSi (left), and (*b*) Energy vs Volume (right).

Total density of states (TDOS) plots estimated by using PBE and mBJ potential are shown in Fig. 2. The band gap values are 0.32 eV and 0.59 eV for PBE and mBJ, respectively. From TDOS plot, non-magnetic insulating characteristics is clearly observed. TB-mBJ is very accurate for the calculations of band gaps of semiconductors and insulators with a Semilocal Exchange-Correlation Potential. Thus, for the further calculations TB-mBJ potential is used. Both valance band (VB) and conduction band (CB) edge have sharp increase in density of states, which suggests that, large number of thermally excited charge carriers are available to contribute in the transport properties. Further, we have calculated the partial density of states (PDOS) to know the contributions from each element. The main contributions to the PDOS in VB region are from $d(t_{2g})$, whereas in CB are from $d(e_g)$ for both Fe and Ru, as they are having the same valance charge. In case of Ti, both VB and CB, the major contributions in PDOS are from the $d(t_{2g})$. For Si, *p*-state contribute the CB. The electronic density of states suggests that, thermally excited electrons from VB have the large density of states available in CB to be occupied and thus contribute in transport property effectively.

We have also calculated the dispersion curve along the high-symmetric k-point **W-L-Γ-X-W**, which is shown in Fig. 3. The maximum of the VB is at Γ point and have the triply degenerate states i.e. band **1**, **2**, and **3** meet at VB maxima. The bottom of the CB is at **X** point and have non-degenerate band **4**. The band structure plot clearly shows the indirect band gap of 0.59 eV. The important feature observed in CB is the appearance of flat band along **Γ-X** direction. This suggest that electrons have the high effective mass [m* = $\hbar/(d^2E/dk^2)$], and thus have the large negative Seebeck coefficient. At **X** point in CB, band **5** is slight above to band **4** i.e. ~100 meV, thus have also accommodate the thermally excited electrons at higher temperature and make contribution efficiently to the transport property.

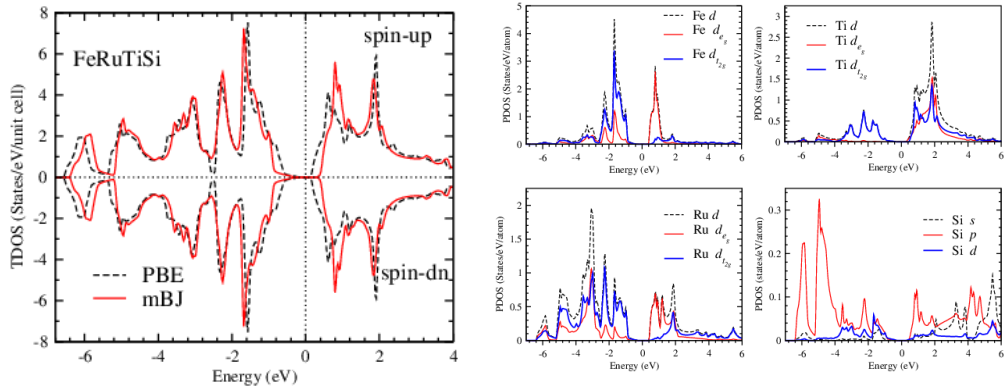

Fig. 2 (left) Total density of states of FeRuTiSi, and (right) partial density of states for Fe, Ru, Ti, and Si.

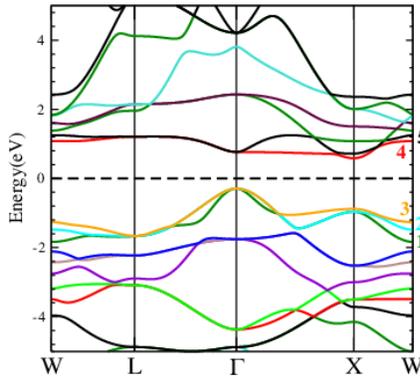

Fig. 3 Dispersion curve for FeRuTiSi along **W-L-Γ-X-W**

Under the constant relaxation time approximation, the calculated transport coefficient using BoltzTraP code is shown in Fig. 4. For the un-doped case we found the large and negative Seebeck coefficient, which is also in consistent with the qualitative prediction from the band structure. With temperature magnitude of *S* decreases as more number of charge carriers get thermally excited, and under the degenerate semiconductor approximation Seebeck coefficient is inversely proportional to the carrier concentration. At 300 K the value of S is found to be -600 μV/K, and remain large at high temperature with ~ -250 μV/K. The value of σ and $\kappa_e$ are calculated by taking the constant relaxation time value, $\tau=10^{-14}$ s, which shows the typical narrow band semiconductor behavior. With temperature both σ and $\kappa_e$ value increase due to increase in carrier concentrations. The calculated $ZT_e$ is found to be ~0.2 in case of intrinsic FeRuTiSi. We have also calculated the transport properties for electron doped case with 1 ×10$^{19}$ /cm$^3$. The corresponding transport coefficients are also shown along with un-doped plot. For the *n*-type dope case, the decrease in magnitude of S is found, which is expected due to increase in carrier concentration. The magnitude still remain large i.e. ~ -140 μV/K. In doped case, S increases with temperature as it is proportional to temperature at fix carrier concentration. The enhancement in electrical conductivity is found due to large carrier concentration and it decrease with temperature due to scattering of charge carriers at high temperature. The corresponding increment in the electronic thermal conductivity is also found. In case of n-type FeRuTiSi, the significant improvement in the calculated value of $ZT_e$ is observed. The maximum value of $ZT_e$ is found to be ~0.86 at 840 K, which is good signature to consider this alloy for *n*-type thermoelectric material with suitable doping.

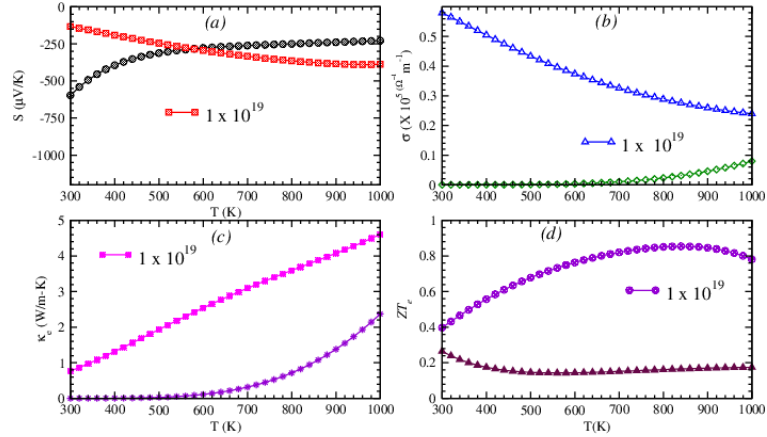

Fig. 4 Seebeck coefficient (*S*), electrical conductivity (σ), electronic thermal conductivity ($\kappa_e$) and figure-of-merit ($ZT_e$) as a function of temperature for FeRuTiSi.

Here, we have calculated the electronic structure and transport properties with mBJ potential. The calculated value of *$ZT_{max}$ (=0.86)* is obtained by taking τ = $10^{-14}$ and by considering only electronic thermal conductivity. In total thermal conductivity, there will be contributions from the lattice part also, which can increase the total thermal conductivity, and thus will have effect on its final *figure-of-merit*. Also, for the insulator, consideration of temperature dependent band gap in transport coefficients calculations are required for better result [11]. The accurate calculation of lattice thermal conductivity demand large strength of computational facilities, which are very costly and also calculations consume more time, thus it is beyond the scope of present work. The formation energy calculation predict that this material can be synthesize [6], therefore it will be interesting to investigate the TE properties of this material experimentally, and rigorous phonon calculations work in near future are highly desirable.

## CONCLUSION

In conclusion, we have investigated the electronic structure and thermoelectric properties of non-magnetic small band gap FeRuTiSi heusler alloy. The band structure calculations give indirect band gap of 0.59 eV. Thermoelectric properties calculated using BoltzTraP code give the maximum $ZT_e$ equal to 0.86 at 840 K for n-type case. This suggest that FeRuTiSi can be a suitable n-type thermoelectric material with appropriate doping and considered for thermoelectric applications in high temperature region.